\documentstyle[12pt]{article}
\begin{document}
\title{A NOTE ON THE MASS OF THE NEUTRINO}
\author{B.G. Sidharth$^*$\\
Centre for Applicable Mathematics \& Computer Sciences\\
B.M. Birla Science Centre, Adarsh Nagar, Hyderabad - 500063 (India)}
\date{}
\maketitle
\begin{abstract}
In the light of the recent Super-Kamiokande experiments which demonstrate
neutrino oscillation, and therefore a non zero mass, it is pointed out how
such a mass has also been deduced theoretically.
\end{abstract}
\footnotetext{E-mail:birlasc@hd1.vsnl.net.in}
Very recent experiments at the Super-Kamiokande Laboratory in Japan have
confirmed results of earlier experiments that there is a neutrino oscillation
and therefore the neutrino has a small mass\cite{r1} in which case the
standard model and problems of dark matter would have to be reexamined, while
other anomalies like the missing solar neutrinos would be explained (cf.also.ref.
\cite{r2}).  The object of the present note is to draw attention to the fact
that a neutrino mass, roughly a billionth that of the electron was
theoretically deduced\cite{r3}.\\
The starting point of this deduction was the model of elementary Fermions
as Kerr-Newman type Black Holes with horizon at the Compton wavelength
in a Quantum Mechanical context which overcomes the naked singularity\cite{r4,r5,r6}.
To put it simply Fermionic or spinorial or doubly connected behaviour results
from the fact that the negative energy solutions predominate within the
Compton wavelength while it is the positive energy solutions that dominate
outside. It was argued that in the case where the mass is vanishingly small
or equivalently the Compton wavelength is very large, we would be encountering
mostly the negative energy solutions, so that not only would the spinorial
behaviour be effected (that is there would be anomalous Bosonization effects),
but also handedness would appear owing to the fact that the negative energy
solutions have opposite parity, as in the case of the neutrino.\\
Infact the Bosonization effects for the neutrino, without any contradiction
with the spin statistics theory can be independently verified\cite{r7}
(cf.Appendix).\\
This fact leads to an interesting consequence. If we used the usual Fermi Dirac
statistics for a neutrino gas under equilibrium conditions, we would
have\cite{r8}
\begin{equation}
PV = \frac{2}{3}U,\label{e1}
\end{equation}
Instead if we take into consideration the Bosonization effect, we would have
\begin{equation}
PV = \frac{1}{3}U,\label{e2}
\end{equation}
In the above $P,V$ and $U$ denote the pressure, volume and energy of the
collection.\\
We also have
$$PV \alpha NkT,$$
where $N$ and $T$ denote the number of particles and temperature
respectively. Interestingly at a given temperature and energy, on comparing
(\ref{e1}) and (\ref{e2}) it can be seen that the number of neutrinos
is effectively halved. Indeed this is exactly the unexplained discrepancy
between the observed number of neutrinos and their predicted number\cite{r9}.\\
If we now consider a fixed number of neutrinos and a given temperature, a
comparison of (\ref{e1}) and (\ref{e2}) shows that the effective energy
$U'$ of the neutrinos would be twice the expected energy $U$. This means
that the neutrino acquires a mass $m$, which can be easily shown to
be, from the above considerations,
\begin{equation}
\frac{mc^2}{k} \approx \sqrt{3}T\label{e3}
\end{equation}
At the present background temperature of about $2^\circ K$, equation
(\ref{e3}) gives a neutrino mass $m,$
\begin{equation}
10^{-9}m_e \le m \le 10^{-8}m_e\label{e4}
\end{equation}
where $m_e$ is the electron rest mass.\\
It is remarkable that (\ref{e3}) is exactly what is required to be deduced
theoretically to justify recent models of lepton conservation or in certain
unification schemes\cite{r2,r10}.\\
We now observe that the balance of the gravitational force
and the Fermi energy of these cold background neutrinos, gives\cite{r11},
$$\frac{GNm^2}{R} = \frac{N^{2/3}\hbar^2}{mR^2},$$
whence,
\begin{equation}
N \sim 10^{90}\label{e5}
\end{equation}
where $N$ is the number of neutrinos, which is correct. (Incidentally we
could verify that the Fermi Temperature $\sim 2^\circ K$, the background
temperature).\\
Incidentally a neutrino mass $m$ could imply an electric charge, by the
following argument: in the case of the electrons it is known that a
fluctuation in the number of electrons $N_e$, viz., $\sqrt{N_e}$ leads to
the following energy balanced equation, keeping in mind the over all
neutrality of electric charge (cf.ref.\cite{r11}):
\begin{equation}
\frac{e^2\sqrt{N_e}}{R} \approx m_ec^2\label{e6}
\end{equation}
which is indeed correct. In the case of the neutrino, if it has a charge $e'$, then a similar
consideration leads to
\begin{equation}
\frac{e'^2 \sqrt{N}}{R} \approx mc^2\label{e7}
\end{equation}
Using values for $m$ and $N$ for the background neutrinos as given in
equations (\ref{e4}) and (\ref{e5}), on comparison of (\ref{e7}) with (\ref{e6}) we get
$$\frac{e'^2}{e^2} \sim 10^{-13}$$
More generally, the right side is given by, $(m/m_e) 10^{-5}$.\\
Interestingly Hayakawa (cf.ref.\cite{r11}) assumes the neutrino mass given
in (\ref{e4}) and there after using considerations similar to those above,
deduces the weak interaction coupling constant also.\\
\vskip 5mm
\noindent {\Large \bf Appendix}
\vskip 5mm
\noindent We can see why, commutators can be used for neutrinos without contradicting
the spin statistics theory\cite{r12}. For Fermionic fields the contradiction
with microscopic causality, which is the real problem arises because the
symmetric propogator, the Lorentz invariant function,
$$\Delta_1(x-x') \equiv \int \frac{d^3k}{(2\pi)^33\omega_k} [e^{-\imath k.(x-x')} +
e^{\imath k.(x-x')}]$$
does not vanish for space like intervals $(x-x')^2 < 0$, where the vacuum
expectation value of the commutator is given by the spectral representation,
$$S_1(x-x') \equiv \imath < 0|[\psi_\alpha (x), \psi_\beta (x')]|0 > = -
\int dM^2[\imath \rho_1(M^2)\Delta_x + \rho_2(M^2)]_{\alpha \beta} \Delta_1
(x-x')$$
Outside the light cone, $r > |t|$, where $r \equiv |\vec x - \vec x'|$ and
$t \equiv |x_o - x_o'|,\Delta_1$ is given by,
$$\Delta_1 (x'-x) = - \frac{1}{2\pi^2 r}\frac{\partial}{\partial r}
K_o(m\sqrt{r^2-t^2}),$$
where the modified Bessel function of the second kind, $K_o$ is given by,
$$K_o(mx) = \int^\infty_o \frac{\cos (xy)}{\sqrt{m^2+y^2}}dy = \frac{1}{2}
\int^\infty_{-\infty} \frac{\cos (xy)}{\sqrt{m^2+y^2}}dy$$
(cf.\cite{r13,r14}). In
our case, $x \equiv \sqrt{r^2-t^2},$ and we have,
$$\Delta_1 (x-x') = const \frac{1}{x} \int^\infty_{-\infty}
\frac{y \sin xy}{\sqrt{m^2+y^2}}dy$$
As we are considering massless neutrinos, going to the limit as $m \to 0$, we
get, $Lt_{m \to 0}\Delta_1 (x-x') = (const.). Lt_{m \to 0}\frac{1}{x}
\int^\infty_{-\infty} \sin xydy = 0$ if $x \ne 0$ (cf. also\cite{r15}).That is the
invariant $\Delta_1$ function vanishes everywhere except on the light cone
$x = 0$, which is exactly what is required.\\
In other words, as pointed out from an alternative standpoint, the neutrino
field can be quantized with commutators, that is there is the anomalous
Bosonic behaviour.\\
Yet another justification for this can be obtained
in\cite{r16}.


\begin{thebibliography}{99}
\bibitem {r1} cf. the website, www.phys.hawaii.edu/$^\sim$ superk for full details.
\bibitem {r2} Feng, D.H.,  He, G.Z. and  Q Li, X., Eds., Proceedings of the
Nankei Summer School on Astrophysics and Neutrino Physics, World Scientific,
Singapore, 1993.
\bibitem {r3} Sidharth, B.G., "Quantum Mechanical Black Holes: Issues and Ramifications",
Proceedings of the International Symposium, 'Frontiers of Fundamental Physics',
Universities Press (Orient Longman), 1998 (In Press). Also xxx.lanl.gov.
quant-ph 9803048.
\bibitem {r4} Sidharth, B.G., "The Universe of Fluctuations". Intl.J.Mod.Phys A
13(15), 1998.
\bibitem {r5} Sidharth, B.G., Gravitation and Cosmology, 4 (2) (14), 1998.
\bibitem {r6} Sidharth, B.G., Ind.J.Pure \& Appd.Phys, \underline{35} (7), 1997.
\bibitem {r7} Sidharth, B.G., "The Bosonic Behaviour of Neutrinos", Technical
Report, BSC-CAMCS 95-04-09.
\bibitem {r8} Huang, K., "Statistical Mechanics", Wiley Eastern, New
Delhi, 1975.
\bibitem {r9} Weinberg, S.,"Gravitation and Cosmology", (John Wiley,
New York, 1972) and several references therein.
\bibitem {r10} Sivaram, C., Astroph. and Space Science, \underline{88},
1982.
\bibitem {r11} Hayakawa, S., Suppl of PTP, 1965.
\bibitem {r12} Bjorken, J.D., and Drell, S.D., "Relativistic Quantum Fields",
Mc-Graw Hill, New York, 1965.
\bibitem {r13} Whittaker, E.T., and Watson, G.N., "A Course of Modern
Analysis", Cambridge University Press, 1962.
\bibitem {r14} Morse, P.M., and  Feschbach, H., "Methods of Theoretical Physics"
(II), Mc-Graw Hill Book Co., New York, 1953.
\bibitem {r15} Davydov, A.S., "Quantum Mechanics", Pergamon Press, Oxford,
1965.
\bibitem {r16} Sallhofer, H.H., "The Maxwell-Dirac Isomorphism", in
"Essays on the Formal Aspects of Electromagnetic Theory", Ed., Lakhtakia, A.,
World Scientific, Singapore, 1993.
\end{thebibliography}
\end{document}